\documentstyle[12pt]{article}

\begin{document}
\thispagestyle{empty}

\begin{center}
{\LARGE\bf ~The Third Family is Different}\\

\vspace{7mm}
\large
Paul H. Frampton

\vspace{5mm}
{\it Institute of Field Physics, Department of Physics
and Astronomy,}\\
{\it University of North Carolina, Chapel Hill,
 NC, 27599--3255, USA}\\
\end{center}

\vspace{1cm}
\begin{abstract}
\noindent
It now appears phenomenologically that the third family of fundamental
fermions may be essentially different fron the first two. Particularly
the high value (174GeV?) of the top quark mass suggests a special role.
In the standard model all three families are treated similarly [becoming
exactly the same at asymptotically high energies] so we need to extend
the model to accommodate the goal of a really different third family.
In this article I describe not one but two such viable extensions, quite
different one from another. The first is the 331 model which predicts
dileptonic gauge bosons. The second is the $Q_6$ model which predicts
additional leptons between 50 and 200 GeV. One expects there are many
other models of this general type characterized by the prediction
of new particles at accessible masses. Supersymmetrization will not be
discussed here.
\end{abstract}
\noindent

In this presentation, I shall address two extensions of the standard
model in which the third family is dealt with asymmetrically with respect to
the lightest two.
First, I shall describe the 331 Model and elaborate on its predictions
including the dilepton,
the neutrino masses and tests in hadronic and leptonic
colliders.
Second, after a review of finite nonabelian groups of order $\leq 31$ and the
question of their gauging and of their chiral anomalies, model building in that
direction is discussed
culminating in a specific model
based on the dicyclic group $Q_6$.

\bigskip
{\bf 331 Model}
\bigskip

Family symmetry is usually taken to mean a horizontal symmetry, either global
or gauged,
under which the three families transform under some non-trivial representation.
The family symmetry is {\it broken} in order to avoid unobserved mass
degeneracies.
In this meaning of family symmetries, there is usually no explanation of {\it
why} there are
three families which are the input. Rather the hope is that the postulated
family
symmetry may explain the observed hierarchies:
\begin{equation}
m_u \ll m_c \ll m_t
\end{equation}
\begin{equation}
m_d \ll m_c \ll m_b
\end{equation}
\begin{equation}
\theta_{13} < \theta_{23} < \theta_{12}
\end{equation}

In the present model, the aim of family symmetry is indeed to attempt to
address such hierarchies {\it and} to explain why there are three
families. This may be a necessary first step to understanding hierarchies?

\bigskip

To introduce the 331 Model\cite{PHF}, the following are motivating factors:

i) Consistency of a gauge theory (unitarity, renormalizability) requires
anomaly cancellation. This requirement almost alone is able to fix all
electric charges and other quantum numbers within one family
of the standard model. This accounts for charge quantization, {\it e.g.} the
neutrality of
the hydrogen atom, without the need for a GUT.

ii)This does not explain why $N_f > 1$ for the number of families but is
sufficiently impressive to suggest that $N_f = 3$ may be explicable by
anomaly cancellation in an extension of the standard model. This requires that
each
extended family have non-vanishing anomaly and that the three families are not
all treated
similarly.

iii) A striking feature of the mass spectrum in the SM is the top mass
suggesting that
the 3rd. family be treated differently and that the anomaly cancellation be
proportional
to: +1 +1 -2 = 0.

iv)There is a " -2 " lurking in the SM in the ratio of the quark electric
charges!

v)The electroweak gauge group extension from $SU(2)$ to $SU(3)$ will add five
gauge bosons. The adjoint of $SU(3)$ breaks into $8 = 3 + (2 + 2) + 1$ under
$SU(2)$.
The $1$ is a $Z^{'}$ and the two doublets are readily identifiable from the
leptonic triplet
or antitriplet $(e^{-}, \nu_e , e^{+})$ as {\it dilepton} gauge bosons
$( Y^{--} , Y^{-})$ with $L = 2$ and $( Y^{++} , Y^{+})$ with $L = -2$.
Such dileptons appeared first in stable-proton GUTs but there the fermions
were non-chiral and one needed to invoke mirror fermions; this is
precisely what is avoided in the 331 Model.  But it is true that the $SU(3)$
of the 331 Model has the same couplings to the {\it leptons} as that of the
leptonic
$SU(3)_l$
subgroup of $SU(15)$ which breaks to $SU(12)_q \times SU(3)_l$  .

\bigskip

Now we are ready to introduce the 331 Model in its technical details: the gauge
group
of the standard model is extended to $SU(3) \times SU(3) \times U(1) $ where
the electroweak
$SU(3)$ contains the standard $SU(2)$ and the weak hypercharge is a mixture of
$ \lambda_8 $
with the $U(1)$. The leptons are in the antitriplet $( e^-, \nu_e, e^+ )_L$ and
similarly
for the $\mu$ and $\tau$.

These antitriplets have $X = 0$ where $X$ is the new $U(1)$ charge. This can be
checked by
noting that the $X$ value is the electric charge of the central member of the
triplet or
antitriplet.

For the first family of quarks we use the triplet $( u, d, D )_L$ with $X =
-1/3$
and the right-handed counterparts in singlets. Similarly, the second family of
quarks is treated. For the third family of quarks, on the other hand, we use
the
{\it antitriplet} $( T, t, b)_L$ with $X = +2/3$. The new exotic quarks D, S,
and T have
charges -4/3, -4/3 and +5/3 respectively.

It is instructive to see how this combination successfully cancels all chiral
anomalies:

The purely color anomaly $(3_L)^3$ cancels because QCD is vector-like.

The anomaly $(3_L)^3$ is non-trivial. Taking, for the moment, arbitrary
numbers $N_c$ of colors and $N_l$ of light neutrinos we find this anomaly
cancels only
if $N_c = N_l = 3$.

The remaining anomalies $(3_c)^2X$, $(3_L)^2X$, $X^3$ and $X(T_{\mu\nu}^2$
also cancel.

Each family separately has non-zero anomaly for $X^3$, $(3_L)^2X$ and
$(3_L)^3$;
in each case, the anomalies cancel proportionally to $+1 +1 -2$ between the
families.

\medskip

To break the symmetry we need several Higgs multiplets. A triplet $\Phi$ with
$X = +1$
and VEV $<\Phi>$ = $(0,0,U)$ breaks 331 to the standard 321 group, and gives
masses to
D, S, and T as well as to the gauge bosons Y and Z'.  The scale U sets the
range of the
new physics and we shall discuss more about its possible value.

The electroweak breaking requires two further triplets $\phi$ and $\phi'$ with
$X = 0$ and $X = -1$ respectively. Their VEVs give masses to d, s, t and to
u, c, b respectively. The first VEV also gives a contribution of an
antisymmetric-in-family type to the charged leptons. To complete a satisfactory
lepton mass matrix necessitates adding a sextet with $X = 0$.

\medskip

What can the scale $U$ be? It turns out that there is not only the lower
bound expected from the constraint of the precision electroweak data,
but also an upper bound coming from a group theoretical constraint within
the theory itself.

The lower bound on $U$ from $Z-Z'$ mixing can be derived from the
diagonalization of the mass matrix and leads to $M(Z') \geq 300GeV$.
The limit from FCNC (the Glashow - Weinberg rule is violated) gives a similar
bound; here the suppression is helped by ubiquitous $(1 - 4sin^2\theta)$
factors.

In these considerations, particularly with regard to FCNC, the special role
played by the third family is crucial; if either of the first two families
is the one treated asymmetrically the FCNC disagree with experiment.

\medskip

The upper bound on $U$ arises because the embedding of the standard $321$
group in $331$ requires that $sin^2\theta \leq 1/4$. When $sin^2\theta = 1/4$,
the $SU(2) \times U(1)$ group embeds entirely in $SU(3)$, and the coupling of
the $X$ charge in principle diverges. Because the phenomenological value is
close to 1/4 -
actually $sin^2\theta (M_Z) = 0.233$ - the scale $U$ must be less than about
$3TeV$ after scaling $sin^2\theta(\mu)$ by the renormalization group. Putting
some
reasonable upper bound on the $X$ coupling leads to an upper bound on the
dilepton
mass, for this 331 Model, of about $800GeV$ [ Here I have allowed one further
Higgs multiplet
- an octet].

\medskip

A very useful experiment for limiting the dilepton mass from below is
polarized muon decay. With the coupling parametrized as $V - \xi A$
where $\xi$ is a Michel parameter, the present limit on $\xi$ is $1 \geq \xi
\geq 0.997$
coming from about $10^8$ examples of the decay. This leads to a lower bound
$M(Y) \geq 300GeV$.

Since
\begin{equation}
(1 - \xi ) \sim (M_W/M_Y)^4
\end{equation}
we deduce that if $(1 - \xi )$ could be measured to an accuracy of $10^{-4}$
the limit would become $M_Y \geq 10 M_W$ and if to an accuracy $10^{-8}$
it would be $M_Y \geq 100M_W$. The first of these is within the realm
of feasability and certainly seems an important experiment to pursue.
The group at the Paul Scherrer Institute near Zurich (Gerber, Fetscher)
is one that is planning this experiment.

\bigskip

{\bf Neutrinos in the 331 Model}

\bigskip

In the minimal 331 Model as described so far, neutrinos are massless and
the model respects lepton number $\Delta L = 0$. Now I shall discuss
soft $L$ breaking for $M(\nu_i) \neq 0$.

Spontaneous breaking of $L$ would lead to a massless (triplet) majoron
in disagreement with experiment. Therefore we consider soft explicit breaking
of $L$.
The lepton families can be written $L_{i\alpha} = (l^-_i, \nu_i, l^+_i )_L$.
Among the Higgs scalars are the $H^{\alpha\beta}$ sextet and the
$\phi^{\alpha}$
triplet. The Yukawa couplings are:

\begin{equation}
h^{ij}_1L^i_{\alpha}L^j_{\beta}H^{\alpha\beta} +
h^{ij}_2L^i_{\alpha}L^j_{\beta}
{\overline{\phi}}_{\gamma}\epsilon^{\alpha\beta\gamma}  + h.c.
\end{equation}
The soft breaking of $L$ is in the triple Higgs couplings:
\begin{equation}
m_1H^{\alpha_1\beta_1}H^{\alpha_2\beta_2}H^{\alpha_3\beta_3}
\epsilon_{\alpha_1\beta_1\gamma_1}\epsilon_{\alpha_2\beta_2\gamma_2} +
m_2(H^{\alpha\beta} {\overline{\phi}}_{\alpha} {\overline{\phi}}_{\beta} +
h.c.)
\end{equation}

The neutrinos acquire mass from one-loop insertions of the soft breaking and
one
finds that provided the VEV $<H^{22}> = 0$ then there is the so-called cubic
see-saw formula:

\begin{equation}
M(\nu_i) = CM(l^-_i)^3/M_W^2
\end{equation}
where $l_i^-$ is the charged lepton corresponding to $\nu_i$ and C is a
constant calculable
in terms of various Yukawa couplings and Higgs masses
but whose absolute value is redundant in the sequel.  As an example, suppose we
adopt the
value for $\nu_{\tau}$ of $29.3eV$, an impressively precise value predicted by
Sciama's cosmology - obviously this is only an example! - then the other
neutrinos
have values $6.2meV$ and $690peV$ (where m is milli- and p is pico-).

\medskip

The $L$ breaking will also contribute to neutrinoless double beta decay but
the rate is around a billion times below present experimental limits.

\medskip

The cubic see-saw with the cube of the charged lepton mass is numerically
quite similar to the more familiar quadratic see-saw with the up quark mass,
but since our present derivation does not involve a right-handed neutrino its
origin is conceptually quite independent. In any case,we can
fit the Hot Dark Matter and MSW requirements but not that for the atmospheric
neutrinos simultaneously just as for the Gell-Mann et al and Yanagida case.

\bigskip

{\bf Phenomenology of the 331 Model}

\bigskip

The dilepton can be produced in a hadron collider such as a $pp$ or $p
\overline{p}$
machine, or in a lepton collider such as $e^+e^-$ or $e^-e^-$.

For the hadron collider the $Y$ may be either pair produced or produced in
association
with an exotic quark [the latter carries $L = \pm 2$]. It turns out that the
associated
production is about one order of magnitude larger. These cross-sections are
calculated in the literature - for a pp collider of the type envisioned there
would
be at least $10^4$ striking events per year.

Surely the most dramatic way to spot a dilepton, however, would be to run a
linear collider in
the $e^-e^-$ mode and find a direct-channel resonance. A narrow spike at
between $300GeV$
and $800GeV$ would have a width at most a few percent of its mass and its decay
to $\mu^-\mu^-$ has no standard model background.

\medskip

{\bf Key Points of 331 Model:}

\bigskip

(i) The family symmetry can attempt to explain the fermion hierarchy and why
there
are three families.

(ii)In the 331 Model, the neutrino mass is either zero or proportional to
the cube of the charged lepton mass, depending on whether or not
one softly breaks $L$.

(iii)The dilepton (300GeV - 800GeV) could produce a narrow resonance in the
$e^-e^-$ mode.

\bigskip

\bigskip

{\bf Finite Groups as Family Symmetries}

\bigskip

As a new topic, let me turn to consideration of generic models of the type
where the symmetry group is $SM \times G$ with $SM$ the standard group
and $G$ is a finite group under which the families tranform under some
non-trivial
representation. This has already been studied for the abelian groups $Z_N$
and for certain non-abelian cases $S_3$ and $S_4$.

Before focusing in on specific groups, let us step back and look at all finite
groups of order
$g \leq 31$. [It is normal to stop at $g = 2^n - 1$ because $g = 2^n$ is always
so rich in groups.]

There are altogether 93 inequivalent such groups: 48 are abelian and the
remaining
45 non-abelian. Groups with $g \geq 32$ might well also be interesting but
surely
lower $g$ is simpler.

\bigskip

{}From any good textbook on finite groups\cite{books} we may find a tabulation
of
the number of finite groups as a function of the order g, the number of
elements in the group.

Amongst finite groups, the non-abelian examples have the advantage
of non-singlet irreducible representations which can be used to inter-relate
families. Which such group to select is based on simplicity: the minimum
order and most economical use of representations\cite{guts}.

Let us first dispense with the abelian groups. These are all made up from
the basic unit $Z_p$, the order p group formed from the $p^{th}$ roots
of unity. It is important to note that the the product $Z_pZ_q$ is identical
to $Z_{pq}$ if and only if p and q have no common prime factor.

If we write the prime factorization of g as:
\begin{equation}
g = \prod_{i}p_i^{k_i}
\end{equation}
where the product is over primes, it follows that the number
$N_a(g)$ of inequivalent abelian groups of order g is given by:
\begin{equation}
N_a(g) = \prod_{k_i}P(k_i)
\end{equation}
where $P(x)$ is the number of unordered partitions of $x$.
For example, for order $g = 144 = 2^43^2$ the value would be
$N_a(144) = P(4)P(2) = 5\times2 = 10$. For $g\leq31$ it is simple
to evaluate $N_a(g)$ by inspection. $N_a(g) = 1$ unless g contains
a nontrivial power ($k_i\geq2$) of a prime. These exceptions are:
$N_a(g = 4,9,12,18,20,25,28) = 2; N_a(8,24,27) = 3$; and $N_a(16) = 5$.
This confirms that:
\begin{equation}
\sum_{g = 1}^{31}N_a(g) = 48
\end{equation}
We shall not consider these abelian cases further, because all their
irreducible representations
are one-dimensional..

\bigskip

Of the nonabelian finite groups, the best known are perhaps the
permutation groups $S_N$ (with $N \geq 3$) of order $N!$
The smallest non-abelian finite group is $S_3$ ($\equiv D_3$),
the symmetry of an equilateral triangle with respect to all
rotations in a three dimensional sense. This group initiates two
infinite series, the $S_N$ and the $D_N$. Both have elementary
geometrical significance since the symmetric permutation group
$S_N$ is the symmetry of the N-plex in N dimensions while the dihedral group
$D_N$ is the symmetry of the planar N-agon in 3 dimensions.
As a family symmetry, the $S_N$ series becomes uninteresting rapidly
as the order and the dimensions of the representions increase. Only $S_3$
and $S_4$ are of any interest as symmetries associated with the particle
spectrum\cite{Pak}, also the order (number of elements) of the $S_N$ groups
grow factorially ($N!$) with N. The order of the dihedral groups increase only
linearly ($2N$) with N and their irreducible representations are all one- and
two- dimensional. This is reminiscent of the representations of the
electroweak $SU(2)_L$ used in Nature.

Each $D_N$ is a subgroup of $O(3)$ and has a counterpart double dihedral
group $Q_{2N}$, of order $4N$, which is a subgroup of the double covering
$SU(2)$ of $O(3)$.

With only the use of $D_N$, $Q_{2N}$, $S_N$ and the tetrahedral group T ( of
order
12, the even permutations subgroup of $S_4$ ) we find 32 of the 45
nonabelian groups up to order 31, either as simple groups or as
products of simple nonabelian groups with abelian groups:
(Note that $D_6 \simeq Z_2 \times D_3, D_{10} \simeq Z_2 \times D_5$ and $
D_{14} \simeq Z_2 \times D_7$ )

$$\begin{tabular}{||c||c||}   \hline
g & \\    \hline
$6$  & $D_3 \equiv S_3$\\  \hline
$8$ & $ D_4 , Q = Q_4 $\\    \hline
$10$& $D_5$\\   \hline
$12$&  $D_6, Q_6, T$ \\ \hline
$14$& $D_7$\\  \hline
$16$& $D_8, Q_8, Z_2 \times D_4, Z_2 \times Q$\\  \hline
$18$& $D_9, Z_3 \times D_3$\\  \hline
$20$& $D_{10}, Q_{10}$ \\  \hline
$22$& $D_{11}$\\  \hline
$24$& $D_{12}, Q_{12}, Z_2 \times D_6, Z_2 \times Q_6, Z_2 \times T$,\\  \hline
 & $Z_3 \times D_4, D_3 \times Q, Z_4 \times D_3, S_4$\\  \hline
$26$& $D_{13}$\\  \hline
$28$& $D_{14}, Q_{14}$ \\  \hline
$30$& $D_{15}, D_5 \times Z_3, D_3 \times Z_5$\\  \hline
\end{tabular}$$
There remain thirteen others formed by twisted products of abelian factors.
Only certain such twistings are permissable, namely (completing all $g \leq 31$
)

$$\begin{tabular}{||c||c||}   \hline
g & \\    \hline
$16$  & $Z_2 \tilde{\times} Z_8$ (two, excluding $D_8$), $Z_4 \tilde{\times}
Z_4, Z_2 \tilde{\times}(Z_2 \times Z_4)$
(two)\\  \hline
$18$ & $Z_2 \tilde{\times} (Z_3 \times Z_3)$\\    \hline
$20$&  $Z_4 \tilde{\times} Z_7$ \\   \hline
$21$&  $Z_3 \tilde{\times} Z_7$ \\    \hline
$24$&  $Z_3 \tilde{\times} Q, Z_3 \tilde{\times} Z_8, Z_3 \tilde{\times} D_4$
\\  \hline
$27$&  $ Z_9 \tilde{\times} Z_3, Z_3 \tilde{\times} (Z_3 \times Z_3)$ \\
\hline
\end{tabular}$$

It can be shown that these thirteen exhaust the classification of {\it all}
inequivalent finite groups up to order thirty-one\cite{books}.

Of the 45 nonabelian groups, the dihedrals ($D_N$) and double dihedrals
($Q_{2N}$), of order 2N and 4N respectively,
form the simplest sequences. In particular, they fall into subgroups of
$O(3)$ and $SU(2)$ respectively,
the two simplest nonabelian continuous groups.

For the $D_N$ and $Q_{2N}$, the multiplication tables, as derivable from the
character tables,
are simple to express in general. $D_N$, for odd N, has two singlet
representations $1,1^{'}$ and $m = (N-1)/2$
doublets $2_{(j)}$ ($1 \leq j \leq m$). The multiplication rules are:

\begin{equation}
1^{'}\times 1^{'} = 1 ; ~~~1^{'}\times 2_{(j)} = 2_{(j)}
\end{equation}
\begin{equation}
2_{(i)}\times 2_{(j)} = \delta_{ij} (1 + 1^{'}) + 2_{(min[i+j,N-i-j])}
+ (1 - \delta_{ij}) 2_{(|i - j|)}
\end{equation}
\noindent

For even N, $D_N$ has four singlets $1, 1^{'},1^{''},1^{'''}$ and $(m - 1)$
doublets
$2_{(j)}$ ($ 1 \leq j \leq m - 1$)where $m = N/2$ with multiplication rules:

\begin{equation}
1^{'}\times 1^{'} = 1^{''} \times 1^{''} = 1^{'''} \times 1^{'''} = 1
\end{equation}
\begin{equation}
1^{'} \times 1^{''} = 1^{'''}; 1^{''} \times 1^{'''} = 1^{'}; 1^{'''} \times
1^{'} = 1^{''}
\end{equation}
\begin{equation}
1^{'}\times 2_{(j)} = 2_{(j)}
\end{equation}
\begin{equation}
1^{''}\times 2_{(j)} = 1^{'''} \times 2_{(j)} = 2_{(m-j)}
\end{equation}
\begin{equation}
2_{(j)} \times 2_{(k)} = 2_{|j-k|} + 2_{(min[j+k,N-j-k])}
\end{equation}

\noindent
(if $k \neq j, (m - j)$)

\begin{equation}
2_{(j)} \times 2_{(j)} = 2 _{(min[2j,N-2j])} + 1 + 1^{'}
\end{equation}

\noindent
(if $j \neq m/2$ )

\begin{equation}
2_{(j)} \times 2_{(m - j)} = 2_{|m - 2j|} + 1^{''} + 1^{'''}
\end{equation}

\noindent
(if $j \neq m/2 $)

\begin{equation}
2_{m/2} \times 2_{m/2} = 1 + 1^{'} + 1^{''} + 1^{'''}
\end{equation}

\noindent
This last is possible only if m is even and hence if N is divisible by {\it
four}.\\

For $Q_{2N}$, there are four singlets $1$,$1^{'}$,$1^{''}$,$1^{'''}$ and
$(N - 1)$ doublets $2_{(j)}$ ($ 1 \leq j \leq (N-1) $).

The singlets have the multiplication rules:

\begin{equation}
1 \times 1 = 1^{'} \times 1^{'} = 1
\end{equation}
\begin{equation}
1^{''} \times 1^{''} = 1^{'''} \times 1^{'''} = 1^{'}
\end{equation}
\begin{equation}
 1^{'} \times 1^{''} = 1^{'''} ; 1^{'''} \times 1^{'} = 1^{''}
\end{equation}

\noindent
for $N = (2k + 1)$ but are identical to those for $D_N$ when N = 2k.

The products involving the $2_{(j)}$ are identical to those given
for $D_N$ (N even) above.

This completes the multiplication rules for 19 of the 45 groups. When needed,
rules for the other groups will be derived.

Since we shall be emphasizing the group $Q_6$, let us be more explicit
concerning it and its destinction
from $D_6$.

 $Q_{2n}$ is defined by the equations:
\begin{equation}
A^{2n} = E
\end{equation}
\begin{equation}
B^2 = A^n
\end{equation}
\begin{equation}
ABA = B
\end{equation}

We can find an explicit matrix representation in the form:

\begin{equation}
A = \left( \begin{array}{cc} cos\theta & sin\theta \\
-sin\theta & cos\theta \end{array} \right)
\end{equation}
with $\theta = \pi/n$. This gives then:
\begin{equation}
A^n = \left( \begin{array}{cc} cos (n\theta) & sin (n\theta) \\
-sin (n\theta) & cos (n\theta) \end{array} \right) \\
= \left( \begin{array}{cc} -1 & 0 \\
0 & -1 \end{array} \right)
\end{equation}
B is chosen as:
\begin{equation}
B = \left( \begin{array}{cc} i & 0 \\
0 & -i \end{array} \right)
\end{equation}
The twelve elements of $Q_6$ are now: $E, A, A^2, A^3, A^4, A^5, B, AB, A^2B,
A^3B, A^4B, A^5B$.

For $D_N$, on the other hand the choice of B is replaced by:
\begin{equation}
B = \left( \begin{array}{cc} 1 & 0 \\
0 & -1 \end{array} \right)
\end{equation}
so that $B^2 = A^n = +1$ instead of $- 1$.

{}From these matrices one can deduce the geometrical interpretation that
whereas $D_6$ is the full
dihedral [{\it i.e.} two-sided] symmetry of a planar hexagon in $O(3)$, $Q_6$
is the full $SU(2)$
symmetry of an equilateral triangle when rotation by $2 \pi$ gives a sign $(-
1)$ and a rotation by
$4 \pi$ is the identity transformation.

\bigskip
\bigskip

{\bf Anomalies and Model Building.}

\bigskip

The models we shall consider have a symmetry comprised of the standard model
gauge group
$SU(3)_C \times SU(2)_L \times U(1)_Y $ producted with a nonabelian finite
group G.

If G is a global (ungauged) symmetry, there are problems if the spacetime
manifold is
topologically nontrivial since it has been shown that any such global
symmetry is broken in the presence of wormholes\cite{global}. From a Local
viewpoint (Local with a capital
means within a flat spacetime neighbourhood) the distinction between a global
and local (gauged) finite symmetry does not exist. The distinction exists only
in a
Global sense (Global meaning pertaining to topological aspects of the
manifold).
In a flat spacetime, gauging a finite group has no meaning. In the
presence of wormholes, expected from the fluctuations
occurring in quantum gravity, gauging G is essential. The mathematical
treatment of such a gauged finite group has a long history\cite{flat}.

In order to gauge the finite group G, the simplest procedure is to gauge
a continuous group H which contains G as a subgroup, and then to spontaneously
break H by choice of a Higgs potential. The symmetry breaking may occur at a
high
energy scale, and then the low energy effective theory will not contain any
gauge potentials or gauge bosons; this effective theory is, as explained above,
Locally identical
to a globally-invariant theory with symmetry G.

For example, consider G = $Q_6$ and H = $SU(2)$. We would
like to use only one irreducible representation $\Phi$ of $Q_6$
in the symmetry-breaking potential $V(\Phi)$. The irreps. of $Q_6$
are $1, 1', 1^{''}, 1^{'''}, 2, 2_S$. The $1^{''}, 1^{'''}$ and
$2_S$ are spinorial and appear in the decompositions only of
$2, 4, 6, 8 ....$ of $SU(2)$. Since $\Phi$ must contain the $1$
of $Q_6$ we must choose from the vectorial irreps. $3, 5, 7, 9 ...$
of $SU(2)$. The appropriate choice is the $7$ represented by a symmetric
traceless
third-rank tensor $\Phi_{ijk}$ with $\Phi_{ikk} = 0$.

For the vacuum expectation value, we choose
\begin{equation}
<\Phi_{111}> = +1; <\Phi_{122}> = -1
\end{equation}
and all other unrelated components vanishing. If we look for the $3\times3$
matrices $R_{ij}$ whicg
leave invariant this VEV we find from choices of indices in
\begin{equation}
R_{il}R_{jm}R_{kn}<\Phi_{lmn}> = <\Phi_{ijk}>
\label{eq18}
\end{equation}
that $R_{31} = R_{32} = 0$ (Use $<\Phi_{3ij}\Phi_{3ij}> = 0$) and that $R_{33}
= \pm1$. Then we find $(R_{11})^3 - 3R_{11}(R_{12})^2 = 1$ (Use $l = m = n = 1$
in $(\ref{eq18})$). This means that if $R_{11} = \cos\theta$ then $\cos3\theta
= 1$ or $\theta_n = 2\pi n/3$. So the elements of $Q_6$ are $A = R_3(\theta_1),
A^2, A^3$ and $B, BA, BA^2$ where $B =$ diag$(i, -i -i)$.

More generally, it can be shown that to obtain $Q_{2N}$ one must use an $N^{\rm
th}$ rank
tensor because one finds for the elements $R_{11}$ and $R_{12}$:
\begin{equation}
\sum_{p=0}^{[N/2]}(-1)^p{N\choose2p} (R_{11})^{N-2p}(R_{12})^{2p} =
\cos N\theta = 1
\end{equation}

If the group H is gauged, it must be free from anomalies. This entails several
conditions which must be met:

(a) The chiral fermions must fall into complete irreducible representations
not only of G but also of H.

(b) These representations must be free of all H anomalies including $(H)^3$,
$(H)^2Y$;
for the cases of H = $O(3), SU(2)$ only the latter anomaly is nontrivial.

(c) If H = $SU(2)$, there must be no global anomaly.

The above three conditions apply to nonabelian H. The case of an abelian H
avoids (a) and (c) but gives rise to additional mixed anomalies in (b).

For nonabelian H, conditions (b) and (c) are straightforward to write down and
solve.
Condition (a) needs more discussion. We shall focus on the special cases
of $O(3) \supset D_N$
and $SU(2) \supset Q_{2N}$.

For $O(3)$ the irreps. are ${\bf 1,3,5,7,....}$ dimensional. $D_N$ has irreps.
(for even $N = 2m$) $1, 1^{'}, 1^{''}, 1^{'''}$ and $2_{(j)} (1 \leq j \leq (m
- 1))$ and these correspond
to:
\begin{equation}
O(3):  {\bf 1}  \rightarrow 1 ; {\bf 3} \rightarrow (1^{'} + 2_{(1)})
\end{equation}

\noindent

The same situation occurs for odd N with irreps. $1, 1^{'}$ and $2_{(j)} ( 1
\leq j \leq (N - 1)/2)$.
If we insist on keeping within the fermions of the standard model, or as close
to that
ideal as possible, nothing beyond a {\bf 3} is necessary because the same
quantum
numbers are not repeated more than three times.

For $SU(2) \supset Q_{2N}$ the corresponding breakdown is:
\begin{equation}
{\bf 1} \rightarrow 1; {\bf 2} \rightarrow 2_{S(1)} ; {\bf 3} \rightarrow 1^{'}
+ 2_{(1)}
\end{equation}
\noindent
where the doublets of $Q_{2N}$, $2_{(1)}$ and $2_{S(1)}$, are defined by Eq.
(20).

These are the principal splittings of a continuous group irrep. into finite
subgroup irreps. we shall
need in our discussions of model building below.

\bigskip

\bigskip

In order to be specific we need to set up a collection of model-building rules.
The main
purpose is to understand why the third family of quarks and leptons is heavy,
and especially why the top
quark is very heavy. Thus we require that:

(A) The t quark mass (and {\it only } the t ) transforms as a {\bf 1} of G.

(B) The b and $\tau$ masses appear as G is broken to $G^{'}$.

We next require that at tree level or one-loop level the second family be
distinguishable from the
first. That is:

(C) After stage (B) first the c mass ($G' \rightarrow G''$), then the
s and $\mu$ masses ($G^{''} \rightarrow
G^{'''}$) are generated.

At stage (C) the u, d and e remain massless.

In addition to the above constraints we require that:

(D) No additional quarks and a minimal number of leptons be introduced beyond
the usual three-family
standard model.

(E) All anomalies are cancelled as described in Section III above, when G is
embedded in the minimal
continuous Lie group H:  $G \subset H$.

We strive to satisfy all of (A) through (E). By the study of specific cases in
the following Section V
we shall see that these constraints are quite nontrivial to satisfy
simultaneously and that
the number of interesting models is small.

\bigskip

\bigskip

{\bf The $Q_6$ Model By Elimination.}

\bigskip

We shall treat special cases for the  nonabelian\cite{ross} group G in turn,
taken from the complete
listing, up to order g = 31, given in Section II above.\\

(a) {\it Dihedral Groups ($D_N$, order g = 2N)}\\

{}From the multiplication rules for $D_N$ we see that if the top quark mass
transforms as a singlet,
as required by rule (A) above, the $t_L$ and $t_R$ must {\it both} be in $1$ or
$1^{'}$ or the
{\it same} $2_{(j)}$. The doublet is unsuitable because it will include a
second quark, violating
rule (A).

To proceed systematically, note that there are three triples of quarks with
common
quantum numbers: (1) $(t,b)_L,(c,s)_L,(u,d)_L$; (2) $t_R,c_R,u_R$ ; and (3)
$b_R,s_R,d_R$.
Since $D_N$ is a subgroup of $SO(3)$ we must look to the rule (E) to see that
each triple must
be in $1+1+1$ or $1^{'}+2_{(1)}$ (equivalent to the vector {\bf 3} of $O(3)$ as
can be deduced
from the $O(3)$ and $D_N$ character tables) of $D_N$ to avoid anomalies. If
$t_L$ and $t_R$
are in $1^{'}$ it follows that $(c,s)_L$, $(u,d)_L$ and $c_R,u_R$ are in
$2_{(1)}$ implying that
charm and up quarks have singlet components in their mass terms, hence
violating rule (A).
If $t_L$ and $t_R$ are in $1$, then so are $(c.s)_L,(u,d)_L$ and $c_R, u_R$
again violating rule (A).

{}From this discussion we deduce that no suitable model based on G = $D_N$
exists.\\

(b) {\it Permutation Groups ($S_N$, order g = N!)}\\

The group $S_3$ is identical to $D_3$ which was excluded in (a) above.
The only other $S_N$ with $g \leq 31$ is $S_4$ which has irreducible
representations
$1, 1^{'}, 2, 3, 3^{'}$. It is a subgroup of $O(3)$ so the triples must be in
$1+1+1$
or $3$ ($5$ of $O(3) \rightarrow 2 + 3^{'}$). Since $3 \times 3 = 1 + 2 + 3 +
3^{'}$ contains
a singlet, neither choice fulfils rule (A).

Hence the groups G = $S_N$ are excluded.\\

(c) {\it Tetrahedral Group (T, order g = 12)}\\

The group T has $1, 1^{'}, 1^{''}, 3$ representations
and $T \subset O(3)$ with irreps. of $O(3)$
decomposing under T as $1 \rightarrow 1$, $3 \rightarrow 3$,
$5 \rightarrow 1^{'} + 1^{''} + 3$.
Thus $t_L$ and $t_R$ must either both be in 1 or 3 and since
$3 \times 3 = 1 + 1^{'} + 1^{''} +2(3)$
both choices violate rule (A).

Hence T cannot be used.\\

(d) {\it Double Dihedral (or Dicyclic) Groups ($Q_{2N}$, order g = 4N)}\\

The above cases (a),(b),(c) are all subgroups of $O(3)$. There exist
counterparts
which are doubled and are subgroups {\it not} of $O(3)$ but of $SU(2)$.

As a first example, consider ${ }^{(d)}D_N = Q_{2N}$ which has representations
$1, 1^{'}, 1^{''}, 1^{'''}$ and $(N - 1)$ doublets $2_{(j)}$
as described above in Section II. In $SU(2)$
the representations decompose under $Q_{2N}$ as: $1 \rightarrow 1, 2
\rightarrow 2_{(2)}$
and $3 \rightarrow 1^{'} + 2_{(1)}$. Thus the possible choices for the three
quark triples are: $1 + 1 + 1$,
$1 + 2_{(2)}$ and $1^{'} + 2_{(1)}$.

We can now go a long way toward fulfilling all the rules in Section IV above.

In the quark sector $t_L$ and $t_R$ must be both $1$ or both $1^{'}$. The
latter leads to other singlet
mass terms from $2_{(1)} \times 2_{(1)}$ and so is excluded. Thus $t_L$ and
$t_R$ must both be $1$.
The simplest choice is then to use $Q_6$ ($Q_{2N}, N \geq 4$ leads to no new
structure) and then to use the
assignments\cite{tom}:\\

$$\begin{array}{ccccccc}

\left( \begin{array}{c} t \\ b \end{array} \right)_{L} &
1 & \begin{array}{c} t_{R}~~~ 1 \\ b_{R} \hspace{0.2in}1^{'} \end{array} &
\left( \begin{array}{c} \nu_{\tau} \\ \tau \end{array} \right)_{L} & 1 &
\tau_{R} & 1^{'} \\

\left. \begin{array}{c} \left( \begin{array}{c} c \\ s \end{array} \right)_{L}
\\
\left( \begin{array}{c} u \\ d \end{array} \right)_{L} \end{array}  \right\} &
2_{S}
 &  \begin{array}{c} \left. \begin{array}{c} c_{R} ~~~ 1\\ u_{R} ~~~ 1
\end{array} \right. \\
\left. \begin{array}{c} s_{R} \\ d_{R} \end{array} \right\} 2 \end{array} &
\left. \begin{array}{c} \left( \begin{array}{c} \nu_{\mu} \\ \mu \end{array}
\right)_{L} \\
\left( \begin{array}{c} \nu_{e} \\ e \end{array} \right)_{L} \end{array}
\right\} & 2_{S}
& \left. \begin{array}{c} \mu_{R} \\ \\ e_{R} \end{array} \right\} & 2

\end{array}$$\\

This choice is free from $(SU(2)^{'})^3$ anomalies. Now consider the mixed
$(SU(2)^{'})^2Y$
anomaly. It is nonvanishing. This is inevitable for any assignment even for
the case of $Q_6$ as can be seen by studying the following table. Normalize the
quadratic Casimir
of the $SU(2)^{'}$ doublet to $+1$ and hence the triplet to $+4$; define
$Q = T_3 + Y$ and the anomaly is:\\

$$\begin{tabular}{||c||c|c|c||}   \hline
 &$2_{S} + 1$ &$ 3$ & $1 + 1 + 1$ \\    \hline\hline
$\left( \begin{array}{c} \nu \\ e^{-} \end{array} \right)_{Li}$ &$ -1$ &$ -4$
&$ 0$   \\
$e^{+}_{Li}$ & $+1$& $+4$ & $0$ \\    \hline
$\left( \begin{array}{c} u \\ d \end{array} \right)_{Li} $& $+1$ & $+4$ & $0$
\\
$\bar{u}_{Li}$ &  $-2$ & $-8$ & $0$ \\
$\bar{d}_{Li}$ & $+1$ &  $+4$ & $0$   \\  \hline
\end{tabular}$$\\

Thus this anomaly adds to $+8$. To cancel this
in the most economial way, can add appropriate additional leptons.

The anomaly $+8$ can be compensated by adding two {\bf 3}s of $SU(2)^{'}$
of left-handed leptons and corresponding singlets of right-handed leptons with
usual quantum
numbers. The quark sector is exactly as in the standard model. The additional
particles might be called Q-leptons. Because their masses break $SU(2)_L$,
their masses should be below about $200$GeV; but phenomenology
dictates that they be above $50$GeV. These additional leptons are the principal
signature
of the $Q_6$ model.

The mass matrices are:\\

$$U = \left( \begin{tabular}{c|c}
$<2_S>$ & $ <2_S> $ \\  \hline
$<1>  $ & $ <1>   $
\end{tabular} \right)$$ \\

and:\\

$$D = L = \left( \begin{tabular}{c|c}
$<1''+1'''+2_S>$ & $ <2_S> $ \\  \hline
$<2>  $ & $ <1'>   $
\end{tabular} \right)$$\\
The way of implementing the hierarchy is by the following steps which comply
with
the rules (A) - (D) of Section IV above:

(A) A VEV to an $SU(2)_L$ doublet which is a singlet of $Q_6$ gives t its heavy
mass without breaking
$Q_6$.

(B) A VEV to a $1^{'}$ gives b and $\tau$ their masses, at the same time
breaking $G = Q_6$ to
$G^{'} = Z_6$.

(C) and (D) The charm quark mass is radiatively generated according to the
diagram of Fig. (1)
[ located at the end of this article.]
which uses a VEV transforming as $(1,2_{S})$ under $SU(2)_L \times SU(2)^{'}$.
Next
the s and $\mu$ acquire their tree-level masses through a $(2, 1^{''} or
1^{'''})$ VEV. These VEVs
break $Z_6$ completely. At this point the u,d and e are still massless.\\

(e) {\it Double Tetrahedral Group (${~~}^{(d)}T$, order g = 24)}\\

The doubled group ${ }^{(d)}T \subset SU(2)^{'}$ has representations $1, 1^{'},
1^{''}, 2_{S}, 2^{'}_{S}, 2^{''}_{S}$ and $3$.
The lowest dimensional representations of $SU(2)^{'}$ decompose as:
$1 \rightarrow 1$,$2 \rightarrow 2_{S}$, $3 \rightarrow 3$.
This leads to a model quite analogous to the $Q_6$ model described above with
the same advantages. Note
that ${ }^{(d)}T$ is isomorphic to $Z_3 \tilde{\times} Q_4$. We assign:

$$\begin{array}{cc}

\left. \begin{array}{c} \left( \begin{array}{c} t \\ b \end{array} \right)_{L}
{}~~~ 1\\
\left. \begin{array}{c} \left( \begin{array}{c} c \\ s \end{array} \right)_{L}
\\
\left( \begin{array}{c} u \\ d \end{array} \right)_{L}  \end{array} \right\}
2_{S} \end{array} \right.&
\left. \begin{array}{c} t_{R}~~~ 1 \\ c_{R} ~~~ 1 \\ u_{R} ~~~ 1 \\
\left.\begin{array}{c}
 b_{R} \\ s_{R} \\ d_{R} \end{array} \right\} 3 \end{array} \right.
\end{array}$$  \\

$$\begin{array}{cc}

\left. \left( \begin{array}{c} \nu_{\tau} \\ \tau \end{array} \right)_{L} ~~~ 1
\right.\\
\left.  \begin{array}{c} \left( \begin{array}{c} \nu_{\mu} \\ \mu \end{array}
\right)_{L}  \\
\left( \begin{array}{c} \nu_{e} \\ e \end{array} \right)_{L} ~~~  \end{array}
\right\}  2_{S}  &
\left. \begin{array}{c} \tau_{R} \\ \mu_{R} \\ e_{R} \end{array} \right\} 3
\end{array} $$   \\

whereupon the mass matrices are:

$$U = \left( \begin{tabular}{c|c}
$<2_S>$ & $ <1> $ \\  \hline
$<2_S>  $ & $ <1>   $
\end{tabular} \right)$$ \\

and:\\

$$D = L = \left( \begin{tabular}{c|c}
$<2_S + 2^{'}_S + 2^{''}_S>$ & $ <3> $ \\  \hline
$<2_S + 2^{'}_S + 2^{''}_S>$ & $ <3> $
\end{tabular} \right)$$\\

To implement the hierarchy complying with rules (A) to (D) of Section IV gives:

(A) A VEV to a $SU(2)_L$ doublet which is a singlet of ${ }^{(d)}T$ gives a
heavy mass to t
without breaking ${ }^{(d)}T$.

(B) A VEV to a {\bf 3} of ${ }^{(d)}T$ gives mass to b and $\tau$.

(C) and (D)  The c quark acquires mass radiatively through
a VEV of $(1 , 2_S )$ via the diagram of Fig. (1). The s and $\mu$ acquire mass
at tree level
through $2^{'}_S$ or $2^{''}_S$ VEVs, breaking $G^{'}$. The u, d and e are
still massless.\\

(f) {\it Other Groups}\\

We have so far fealt with 22 of the 45 nonabelian groups with $g \leq 31$.
Another 11 are not
simple so fall outside our search. The remaining 12 are twisted products of
$Z_N$'s and of
orders: g = 16 (5), 18, 20, 21, 24 (2) and 27 (2). Note that one of these has
already been
discussed at length since ${ }^{(d)}T = Z_3 \tilde{\times} Q_4$.

Of the remainder, $Z_9 \tilde{\times} Z_3 = \Delta(27)$, a subgroup of $SU(3)$.
None of the others is embeddable in an $SU(2)^{'}$, since all such groups are
considered in (a) -(e)
above. The only other one which embeds in $SU(3)$ is $\Delta(24) = T \times
Z_2$
but triplets do not allow a hierarchy following our rules of Section IV. The
only other Lie
group of interest with irreducible representations $\leq 3$ is $SO(4) = SU(2)
\times SU(2)$
with a simple direct product. None of our list embeds minimally in $SO(4)$
because their
products are twisted, and not simple.

Thus only two possibilities - $Q_{6}$ [the simpler choice]and ${ }^{(d)}T$
- permit a fermion hierarchy of the type we have specified.\\

{\bf General Comments On $Q_6$ Model.}\\

We have exhibited a model with a nonabelian discrete flavor group $Q_6$ where
$Q_6$ can be embedded in an anomaly-free
$SU(2)$. The top quark can be much heavier than all other quark flavors because
it alone has a $Q_6$-invariant mass.

The breaking of $Q_6$ gives rise sequentially to the other fermion masses:
first b and $\tau$; then c; and finally
s and $\mu$. The first family masses are so tiny that they are neglected at the
order considered here.
A testable prediction is the occurrence of the  additional leptons
in the mass range $50$GeV to $200$GeV.

We should mention the point that the spontaneous breaking of discrete
symmetries
always gives rise to the danger of unacceptable cosmological
domain walls\cite{zel}. However this danger can always be avoided  in the $Q_6$
model
by adding explicit soft breaking in the potential function.

Many outstanding questions remain such as: Can this scheme be made more
quantitative?
How does $Q_6$ arise in a more complete framework\cite{SU3}?

Although we have removed the extreme hierarchy of the Yukawa couplings, it has
been replaced by a
more involved Higgs sector. This seems inevitable; the point is that the
discrete group $Q_6$ leads
to a new viewpoint that provides a first step to understanding the mass
spectrum of quarks and leptons,
particularly why the top quark mass is so different from all other fermion
masses.\\

\bigskip

\bigskip

{\bf Comparison of $331$ and $Q_6$ Models.}\\

\bigskip

We have explored the premise that the third family is different from the first
two even at asymptotically high energy. This premise could be false but is
suggested
by the fermion mass hierarchy. We have given two quite different
implementations of this
premise.

Both models exploit anomaly cancellation in an important way. The $331$ Model
has the advantage
that a first step is made to explaining why there are three families; the key
prediction is
the dilepton. The $Q_6$ model takes as given the three-family structure and has
the advantage of
singling out the top quark as unique among the fermion masses; here the
Q-leptons are predicted.

Both models deserve further investigation, including the question of whether
motivations and tests can be devised which can further constrain the
possibilities.

\bigskip

\bigskip

{\bf Acknowledgements.}

I wish to thank my coauthors listed in References 1 and 10.
This work was supported in part by the U.S. Department of
Energy under Grants DE-FG05-85ER-40219, Task B.

\newpage

\begin{figure}[h]

\vspace*{2.0cm}

\setlength{\unitlength}{1.0cm}

\begin{picture}(15,12)

\thicklines

\put(2,6){\thicklines\vector(1,0){1}}
\put(3,6){\line(1,0){1}}

\multiput(4,6)(0.6,0.6){5}{\line(1,1){0.4}}

\put(7,3){\vector(-1,1){2}}
\put(5,5){\line(-1,1){1}}

\multiput(7,9)(0,0.4){5}{\line(0,1){0.2}}

\put(12,6){\vector(-1,0){1}}
\put(11,6){\line(-1,0){1}}

\multiput(10,6)(-0.6,0.6){5}{\line(-1,1){0.4}}

\multiput(7,3)(0,-0.4){5}{\line(0,-1){0.2}}

\put(7,3){\vector(1,1){2}}
\put(9,5){\line(1,1){1}}

\put(8.3,7.7){\vector(1,-1){0.4}}

\put(5.7,7.7){\vector(-1,-1){0.4}}

\put(7,9.8){\vector(0,1){0.4}}

\put(7,2.2){\vector(0,-1){0.4}}

\put(6.854,1.127){${\bf \times}$}

\put(6.854,10.75){$\times$}

\put(6.2,12){
$ \left \langle \ {(1,2_S)} \ \right \rangle_{\ 0} $}

\put(6.2,0.2){
$ \left \langle \ {(2,1)} \ \right \rangle_{\ 0} $}

\put(0.2,6){$ c_L (2,2_S) $}

\put(12.7,6){$ {c_R (1,1)} $}

\put(7,6){\oval(1,1)[t]}
\put(7,6){\oval(1,1)[br]}
\put(7,5.5){\vector(-1,0){0.2}}

\put(9,8){$(2,1)  $}
\put(9,4){$ t_L (2,1) $}

\put(4.5,8){$ (2,2_S) $}
\put(4.5,4){$ t_R (1,1)$}

\end{picture}

\caption{One loop diagram contributing to the charm quark mass.}

\label{Fig. 1}

\end{figure}
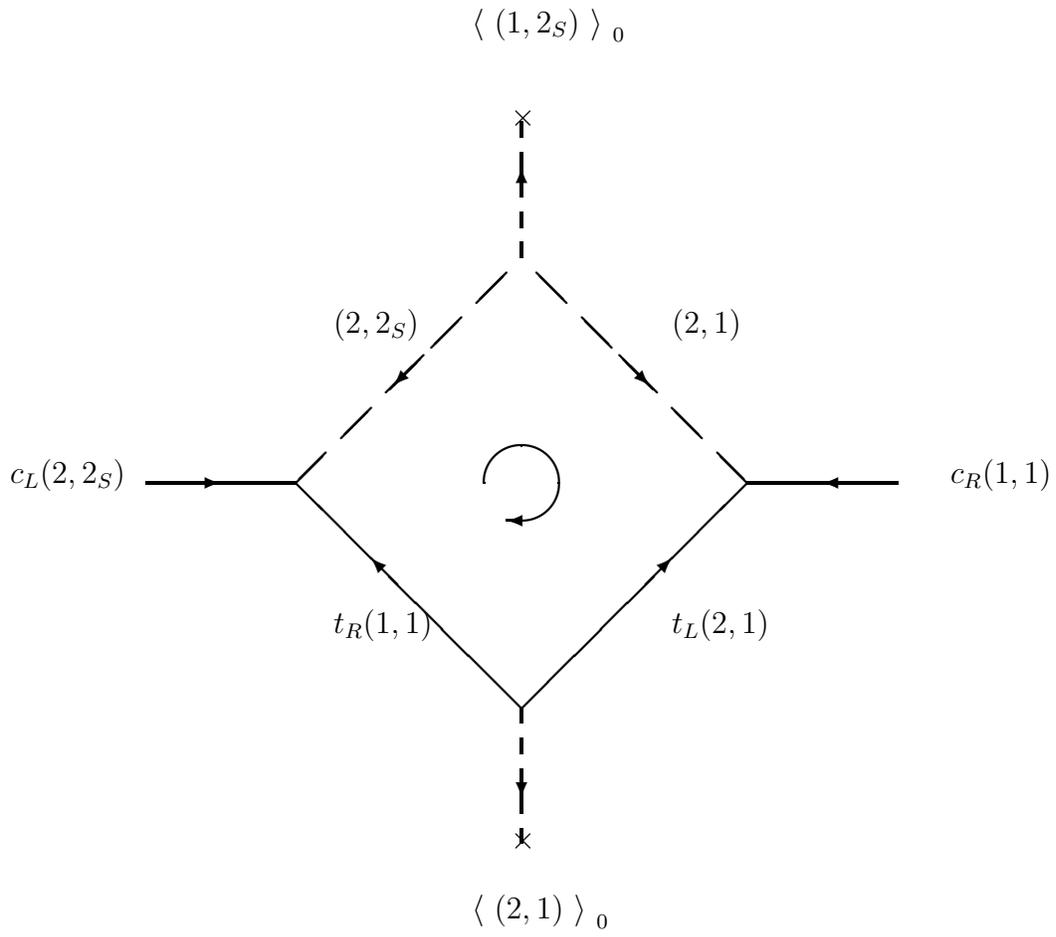


\begin{thebibliography}{99}
\bibitem{guts}
Another route to discrete family
symmetry is through the previous $SU(N)$ family symmetry models\cite{SUN}
. Here one can consider the breaking pattern
\begin{equation}
SU(N)\rightarrow SU(5)\times SU(5-N)\times U(1)\rightarrow SU(5)\times G
\end{equation}
where G is a discrete group. Such models typically have more than just
three complete families until $G$ is broken.
\bibitem{rootN}
Note that even though N 3-vector VEVs are needed to break $SO(3)$ to $D_N$
the top Yukawa has a factor $N^{-1/2}$ suppression as can be seen by
comparing the t and W masses.
\end{thebibliography}
\end{document}